\documentclass[manuscript]{aastex61}

\accepted{\today}

\shorttitle{First detection of [$^{13}$C I] $^{3}${\it{P}}$_{1}$--$^{3}${\it{P}}$_{0}$ emission in 49 Ceti}
\shortauthors{Higuchi et al.}

\begin{document}
\title{First detection of submillimeter-wave [$^{13}$C I] $^{3}${\it{P}}$_{1}$--$^{3}${\it{P}}$_{0}$ emission in a gaseous debris disk of 49 Ceti with ALMA}

\correspondingauthor{Aya E. Higuchi}
\email{aya.higuchi@nao.ac.jp}

\author[0000-0002-9221-2910]{Aya E. Higuchi}
\affil{National Astronomical Observatory of Japan, Osawa, Mitaka, Tokyo 181-8588, Japan}

\author{Yoko Oya}
\affiliation{Department of Physics, The University of Tokyo, Hongo, Bunkyo-ku, Tokyo 113-0033, Japan}
\affiliation{Research Center for the Early Universe, The University of Tokyo, 7-3-1, Hongo, Bunkyo-ku, Tokyo 113-0033, Japan} 

\author{Satoshi Yamamoto}
\affiliation{Department of Physics, The University of Tokyo, Hongo, Bunkyo-ku, Tokyo 113-0033, Japan}
\affiliation{Research Center for the Early Universe, The University of Tokyo, 7-3-1, Hongo, Bunkyo-ku, Tokyo 113-0033, Japan}

\begin{abstract}

We have detected the submillimeter-wave fine-structure transition ($^{3}${\it{P}}$_{1}$--$^{3}${\it{P}}$_{0}$) of $^{13}$C, [$^{13}$C I],
in the gaseous debris disk of 49 Ceti with the Atacama Large Millimeter/submillimeter Array (ALMA).
Recently, the [C I] $^{3}${\it{P}}$_{1}$--$^{3}${\it{P}}$_{0}$ emission has been spatially resolved in this source with ALMA.
In this dataset, the $F$=3/2--1/2 hyperfine component of [$^{13}$C I], which is blue-shifted by 2.2 km~s$^{-1}$ from the normal species line, 
[C I], has been identified in the outer part of the 49~Ceti disk, thanks to the narrow velocity widths of the gas components.
The [C I]/[$^{13}$C I] line intensity ratio is found to be 12$\pm$3, 
which is significantly lower than the $^{12}$C/$^{13}$C abundance ratio of 77 in the interstellar medium.
This result clearly reveals that the [C I] emission is optically thick in 49 Ceti at least partly, as speculated by the previous work.
As far as we know, this is the first detection of [$^{13}$C I] $^{3}${\it{P}}$_{1}$--$^{3}${\it{P}}$_{0}$ emission at 492~GHz 
not only in debris disks but also in the interstellar medium.

\end{abstract}

\section{Introduction} 

49 Ceti is known as a famous nearby gaseous debris disk source (40 -- 50~Myr; \cite{zuc12, zuc19}), whose distance is 57$\pm$0.3~pc \citep{gai18}.
Toward this source, emission of CO \citep{zuc95,den05,hug08,hug17} and its isotopologues \citep{moo19}, 
far-ultraviolet absorption of [C I], far-infrared [C II] emission \citep{rob14}, 
and submillimeter-wave [C I] emission \citep{hig17} have been reported so far.
Very recently, \cite{hig19} presented a high-resolution image of the submillimeter-wave 
[C I] $^{3}${\it{P}}$_{1}$--$^{3}${\it{P}}$_{0}$ emission toward this source with a high signal-to-noise (S/N) ratio by using 
the Atacama Large Millimeter/submillimeter Array (ALMA).
Their result shows that the [C I] emission is quite bright in the gas component around 49 Ceti.

Although the spatial distributions of the CO and [C I] emissions are slightly different from each other, 
the CO and [C I] emissions are both bright at the distance of 30 to 70~au from the central star along the major axis of the disk.
It seems likely that these emissions are optically thick in some parts of the disk \citep{kra19, hig19}.
For this reason, the column densities and excitation temperatures have not been derived precisely.
As the result, the C/CO abundance ratio, which reflects the effect of photodissociation by the stellar and/or interstellar UV radiation, 
has not been discussed quantitatively.
Observations of the isotopic species of C and CO are thus crucial to step forward toward characterizing the gas component and 
understanding its origin.

The rest frequencies of the $^{3}${\it{P}}$_{1}$--$^{3}${\it{P}}$_{0}$ transitions of [C I] and [$^{13}$C I] at 492 GHz 
were precisely measured by the laboratory microwave spectroscopy \citep{yam91}. 
The transition of the [$^{13}$C I] emission consists of two hyperfine components, $F$=3/2--1/2 and $F$=1/2--1/2, whose rest frequencies are
blue- and red-shifted by 2.2~km~s$^{-1}$ and 0.3~km~s$^{-1}$, respectively, from that of the normal species.
Although the $F$=1/2--1/2 component is hardly separated from the normal species line, the $F$=3/2--1/2 component
can be observed toward narrow-line sources with high velocity resolution observations.
Since the [C I] emission in 49 Ceti is bright and narrow in the outer part of the disk (120 to 180~au),
we carefully inspected their spectra to identify [$^{13}$C I].
In this letter, we report the first detection the [$^{13}$C I] emission in 49 Ceti.

\section{Dataset}

We used the ALMA data of the [C I] $^{3}${\it{P}}$_{1}$--$^{3}${\it{P}}$_{0}$ observations toward the gaseous debris disk of 49 Ceti, 
whose original purpose was to understand the origin of the gas component.
Details of the [C I] observations (weather condition, calibrators, spatial and velocity resolutions, and data qualities) 
have already been reported by \cite{hig19}.
The parameters of the [C I] $^{3}${\it{P}}$_{1}$--$^{3}${\it{P}}$_{0}$ and [$^{13}$C I] $^{3}${\it{P}}$_{1}$--$^{3}${\it{P}}$_{0}$ lines
are listed in Table 1. 

Data reduction was performed by using version 5.5.0 of the Common Astronomy Software Applications (CASA) package \citep{mcm07}. 
In order to identify the faint emission, Briggs weighting of +0.5 was used to obtain the spectra \citep{pet12}.

\section{Results and Discussions}

\subsection{[C I] $^{3}${\it{P}}$_{1}$--$^{3}${\it{P}}$_{0}$ line emission}

\begin{figure}
\epsscale{1.1}
\plotone{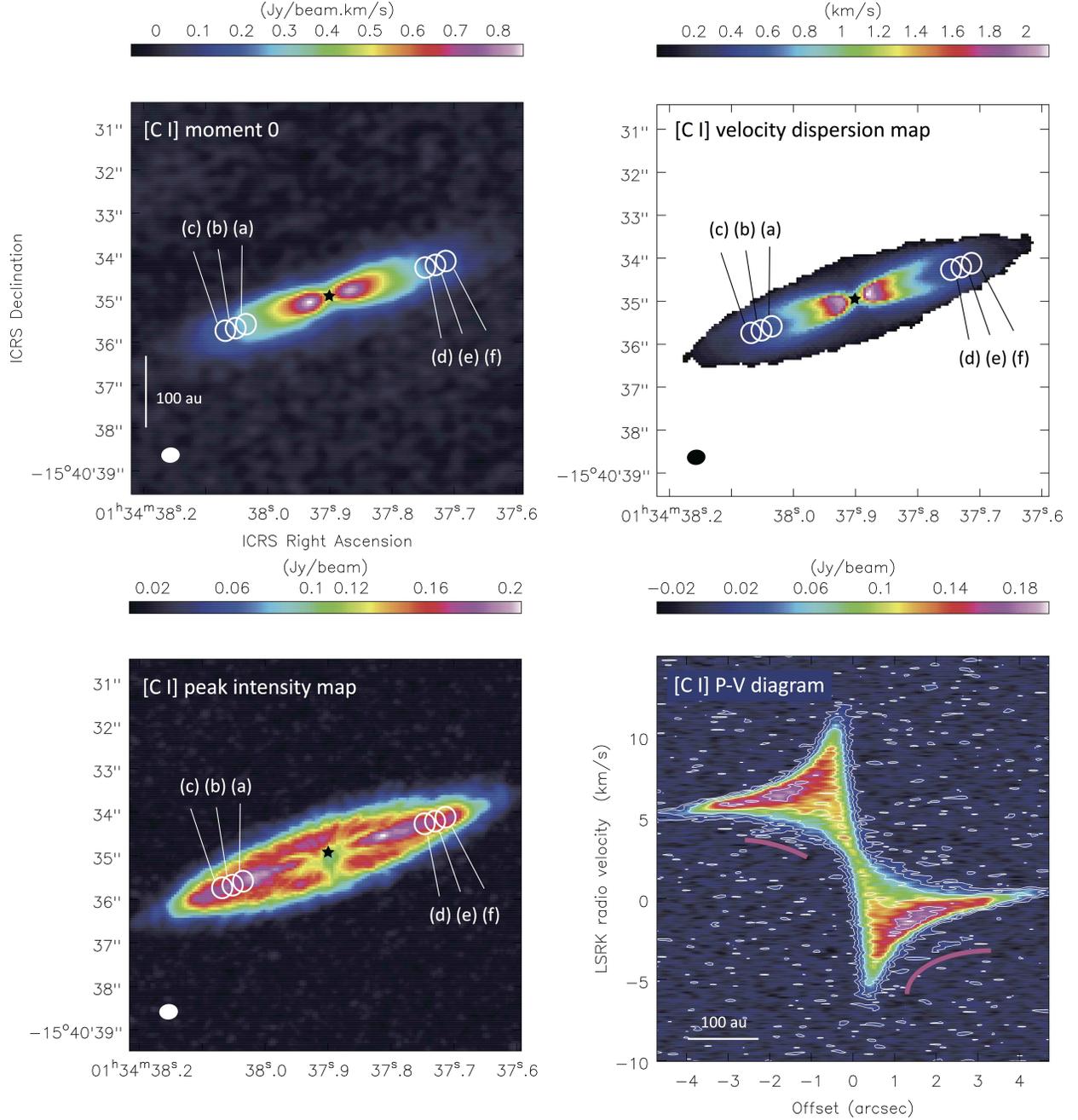}
\caption{
[Top-left] The integrated intensity map of the [C I] emission. A star mark indicates the position of the central star.
Open circles (a) - (f) indicate the positions of individual spectra shown in Figure \ref{fig2}.
[Top-right] The velocity dispersion map of the [C I] emission.
[Bottom-left] The peak intensity map of the [C I] emission. Open circles (a) - (f) indicate the positions of individual spectra shown in Figure \ref{fig2}.
[Bottom-right] The position-velocity diagram of the [C I] emission along the major axis prepared by using the CASA {\it impv}. 
The starting position is ($\alpha_{J2000}$, $\delta_{J2000}$) = 
(01$^{\mbox h}$34$^{\mbox m}$38$^{\mbox s}.$210, ~$-$~$\!\!$15$^{\circ}$40$^{\arcmin}$36$\arcsec.$450),
while the end position is ($\alpha_{J2000}$, $\delta_{J2000}$) = 
(01$^{\mbox h}$34$^{\mbox m}$37$^{\mbox s}.$594, ~$-$~$\!\!$15$^{\circ}$40$^{\arcmin}$33$\arcsec.$450)
with P.A.= $-$ 72$^{\circ}$. 
Contours are 2~$\sigma$, 5~$\sigma$, 10~$\sigma$, 20~$\sigma$, 30~$\sigma$ levels (1~$\sigma$ = 5.5~mJy~beam$^{-1}$).
Pink lines guide a marginal emission feature of [$^{13}$C I] seen in the velocity range of $-$3 to $-$5~km~s$^{-1}$ and that of +3 to +5~km~s$^{-1}$ 
(see text).}
\label{fig1}
\end{figure}

The top left panel of Figure \ref{fig1} shows the [C I] integrated intensity map.
The velocity range for integration is from $-$6 to 11.5~km~s$^{-1}$, where the systemic velocity is 2.8~km~s$^{-1}$.
Details of its spatial distribution have already been reported by \cite{hig19}.
Although a double-peak distribution centered at the star position is seen in the integrated intensity map at 
a 30 au scale in radius, these two peaks do not correspond to the actual brightness peaks of the [C I] emission.
This is because the integrated intensity is largely affected by the velocity width due to the Keplerian motion.
The top right panel of Figure \ref{fig1} shows the velocity dispersion map of the [C I] emission, supporting the effect described above.

The bottom left panel of Figure \ref{fig1} shows the [C I] peak intensity map.
This map indicates that the region with the bright [C I] emission extends from 30 to 180~au along the major axis. 
The bottom right panel of Figure \ref{fig1} shows the position-velocity (P-V) diagram of the [C I] emission along the major axis prepared 
by using the CASA {\it impv}, which has been presented by \cite{hig19}.
It is confirmed that the region with bright the [C I] emission extends to the outer part of the disk.
Moreover, the line width tends to be narrower in the outer part.
Based on the peak intensity map and the P-V diagram, we thought that the $F$=3/2--1/2 component of 
[$^{13}$C I] could be found in the outer part.

The intensity of the $F$=3/2--1/2 component of [$^{13}$C I] is estimated to be 1/116 of the [C I] intensity, if the $^{12}$C/$^{13}$C 
abundance ratio is that of the interstellar medium (77) \citep{wil94} and the [C I] line is optically thin.
Here, the effect of the hyperfine splitting is considered.
Since the S/N ratio of the P-V diagram of [C I] is about 30, it is difficult to identify such a faint line.
However, we may be able to find the [$^{13}$C I] line, if the [C I] line is optically thick.
With this in mind, we prepared the P-V diagram with contours from 2~$\sigma$ to 30~$\sigma$
(1$\sigma$ = 5.5~mJy~beam$^{-1}$).

A marginal emission feature is seen in the region where the velocity of $-$3 to $-$5~km~s$^{-1}$ (northwest) 
and that of $+$3 to $+$5~km~s$^{-1}$ (southeast), as guided by the pink lines.
We thought that this feature might be a hint of the [$^{13}$C I] emission.

\subsection{Detection of [$^{13}$C I] $^{3}${\it{P}}$_{1}$--$^{3}${\it{P}}$_{0}$ line emission}

\begin{figure}
\epsscale{1.2}
\plotone{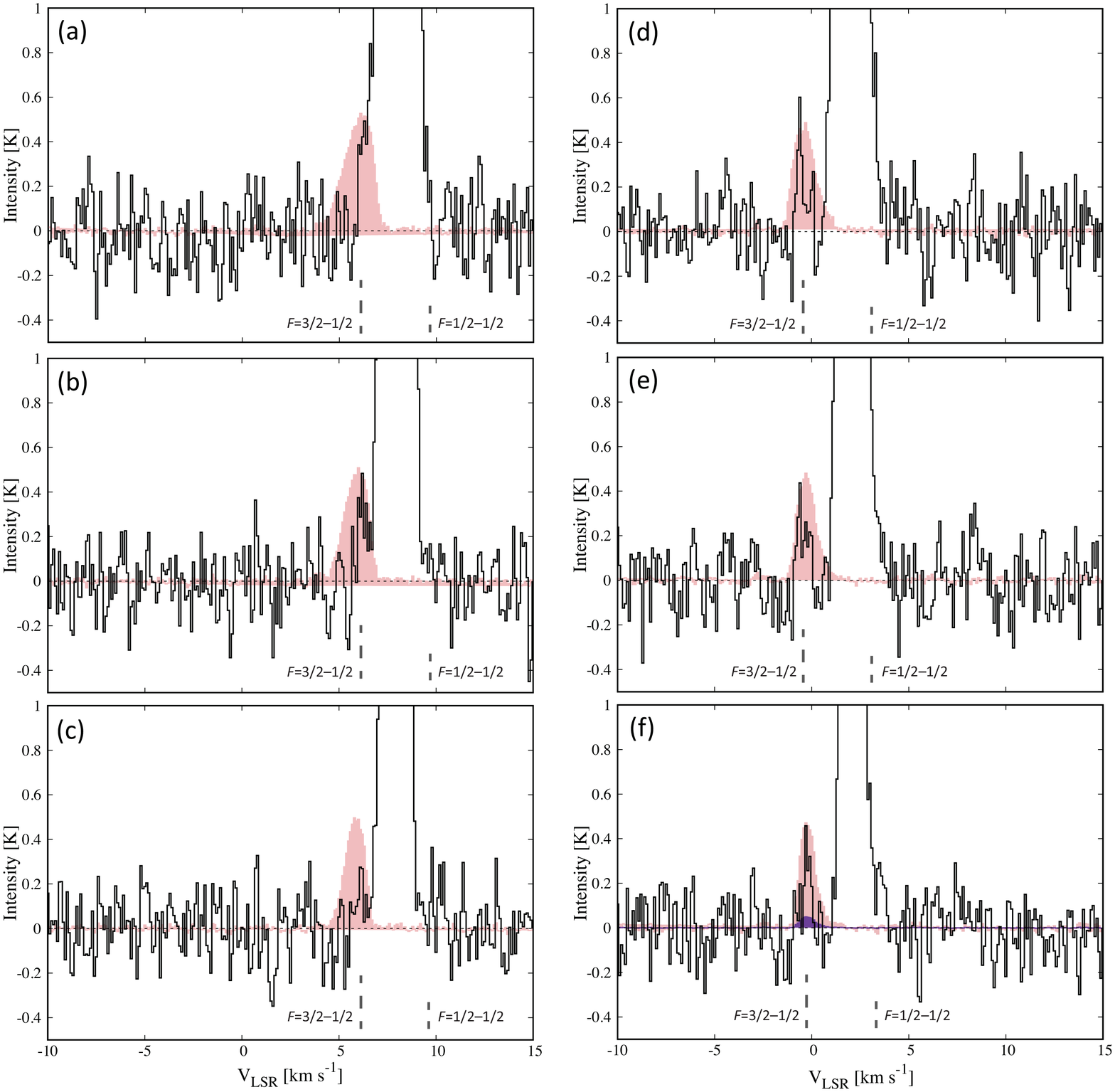}
\caption{The $^{3}${\it{P}}$_{1}$--$^{3}${\it{P}}$_{0}$ transitions of the [C I] and [$^{13}$C I] spectra.
The overlaid spectra with pink color show the [C I] spectra whose intensity is scaled by 1/12 and whose velocity is blue-shifted by 2.2~km~s$^{-1}$.
For the panel (f), the overlaid spectrum with purple color shows the [C I] spectrum whose intensity is scaled by 1/116 (see Section 3.1) 
and whose velocity is blue-shifted by 2.2~km~s$^{-1}$.
The horizontal axes represent $V_{\rm{LSR}}$ of the [C I] emission.}
\label{fig2}
\end{figure}

To investigate the marginal feature in more detail, we prepared the spectra toward the six circular areas indicated in Figure \ref{fig1}.
The results are shown in Figure \ref{fig2}.
These spectra are ones averaged over the 0.$\arcsec$5 areas in diameter which correspond to the spatial resolution.
Although the observed areas are overlapped with each other by a half beam, they provide independent information according to the 
Nyquist sampling theory \citep{wil09}.

In Figure \ref{fig2}, the dashed lines indicate the expected velocities of two hyperfine components of [$^{13}$C I].
A faint spectral feature appears just at the expected velocity of the $F$=3/2--1/2 component of the [$^{13}$C I] line
in the spectra of the northwestern positions (Figures \ref{fig2} d - f).
The overlaid spectra with pink color show the [C I] spectra whose intensity is scaled by 1/12 and whose 
velocity is blue-shifted by 2.2~km~s$~^{-1}$ (the velocity difference between the $F$=3/2--1/2 line of [$^{13}$C I] and the [C I] line).
The overlaid spectra well correspond to the faint features.
From these results, we concluded that the faint feature is the $F$=3/2--1/2 component of the [$^{13}$C I] line.

Since the root-mean-square (rms) noise level of the spectra is 0.11~K at a resolution of 0.1 km~s$^{-1}$,
the [C I] lines are detected in the northeastern positions of (d), (e), and (f) of Figures \ref{fig1} with the 
3 - 5~$\sigma$ confidence level (see Table 2).
For the southeastern positions, we cannot detect the [$^{13}$C I] line definitively, 
because of the heavy contamination of the wing component of the [C I] emission.
Nevertheless the overlaid spectra in pink color could be consistent with the observed spectra even for Figures \ref{fig2} a - c. 


\subsection{[C I] and [$^{13}$C I] intensity ratio}

In the above analysis, we derived the peak intensity ratio of [C I] relative to [$^{13}$C I] to be 12$\pm$3. 
Considering the $^{12}$C/$^{13}$C abundance ratio in the interstellar medium of 77 \citep{wil94},
the observed [C I]/[$^{13}$C I] intensity ratio is much lower than the elemental ratio.
This result shows that the [C I] emission is optically thick.
The peak optical depth ($\tau$) of the [C I] emission is roughly estimated to be 10$\pm$3 by considering the relative 
intensities of the hyperfine components.
Here, we assumed the excitation temperature of [$^{13}$C I] is identical to that of [C I].
Since the optical depth that of [C I] is higher than that of [$^{13}$C I],
the excitation temperature of [C I] could be higher than that of [$^{13}$C I] in principle.
In this case, the peak optical depth measured above is a lower limit, 
and hence, our conclusion of optically thick [C I] does not change.

The spectral shape is apparently different between the [C I] and [$^{13}$C I] spectra: the velocity width is narrower 
in [$^{13}$C I] than in [C I].
This can be understood as the optical depth effect at the peak velocity: the optical depth at the intensity peak of the 
[C I] spectrum is obviously thick, 
while at other velocities of the [C I] that in the rest part of the [C I] emission could be lower.

Here, we evaluated the column density, $N$(C), and the gas kinetic temperature, $T{\rm_{K}}$, by using 
the non-LTE code (RADEX)\footnote{http://var.sron.nl/radex/radex.php}. 
We optimized these two parameters to reproduce the [C I] intensity and its optical depth (10$\pm$3) derived above.
The major collision partner is not well defined for this situation, so that we simply assume the H$_{2}$ density of 10$^{4}$~cm$^{-3}$ 
(critical density of [C I]) for convenience in the calculation. Here, the velocity width is assumed to be 1~km~s$^{-1}$.
Although the contribution of electrons is considered for the CO excitation \citep{kra19}, we ignored it for [C I] in this calculation.
As the result, we obtained the following parameters: $T{\rm_{K}}$ = 18$\pm$7~K and $N$(C) = (1.5$\pm$1.0) $\times$ 10$^{18}$~cm$^{-2}$.
The excitation temperature is derived to be 18$\pm$7~K, being comparable to $T{\rm_{K}}$.
This indicates that the lines are thermalized.
The column density is by an order of magnitude higher than the value derived from the single dish observations by \cite{hig17}.

\subsection{A caveat for the evaluation of the optical depth}

In the previous section, we assumed the $^{12}$C/$^{13}$C elemental abundance ratio of 77 in order to evaluate the optical depth. 
However, we need to discuss its validity in the debris disk condition, which would be affected by stellar and/or interstellar UV radiation. 
If the atomic carbon (C) or the ionized carbon (C$^{+}$) 
is the major reservoir of gaseous carbon, the $^{12}$C/$^{13}$C ratio of the atomic carbon should be close to 77. 
This is because the C$^{+}$/C ratio is determined by the ionization process, 
which is free from the isotope fractionation. 
On the other hand, the $^{12}$C/$^{13}$C ratio of the atomic carbon could be higher than 77, 
if the major reservoir of gaseous carbon is CO. In this case, $^{13}$C$^{+}$ formed by photodissociation can react with the abundant $^{12}$CO as \citep{lan84}: 
\begin{equation}
{^{13}}\rm{C}^{+} + {^{12}}\rm{CO} \rightarrow {^{12}}\rm{C}^{+} + {^{13}}\rm{CO} + 35~\rm{K}.
\end{equation}
For this isotope exchange reaction, $^{13}$C$^{+}$ and $^{13}$C become deficient \citep{lan84}.
This mechanism particularly works in cold conditions, where the backward reaction of the equation (1) is not effective.
Later chemical model simulations indeed confirmed this effect \citep{fur11,rol13}. 
If the $^{12}$C/$^{13}$C ratio is higher than 77 in 49 Ceti due to this mechanism, the [C I] optical depth evaluated in this study 
from the [C I] and [$^{13}$C I] intensities would be higher than 10. 
For confirmation, observations of the $^{3}${\it{P}}$_{2}$--$^{3}${\it{P}}$_{1}$ (809~GHz) lines of [C I] and [$^{13}$C I] are helpful,
which can be done with ALMA.

\subsection{Future perspective}

Although observations of [$^{13}$C I] are essential for the accurate evaluation of the column density of atomic carbon,
they have been very sparse so far even in the interstellar medium.
This is because the [C I] line is believed not to be optically thick.
For the $^{3}${\it{P}}$_{2}$--$^{3}${\it{P}}$_{1}$ transition of [$^{13}$C I] (809~GHz), the frequency is shifted from [C I] by 152~MHz 
(56~km~s$^{-1}$; \cite{kle98}), providing a chance to detect the [$^{13}$C I] line emissions.
In fact, the first detection of the [$^{13}$C I] $^{3}${\it{P}}$_{2}$--$^{3}${\it{P}}$_{1}$ line emission in Orion IRc2 was 
achieved by \cite{kee98} and further observations have been reported by \cite{tie01}.

On the other hand, it is difficult to detect the $^{3}${\it{P}}$_{1}$--$^{3}${\it{P}}$_{0}$ transition of [$^{13}$C I] (492~GHz)
because its frequency is shifted from the [C I] line emissions only by 3.6~MHz (2.2~km~s$^{-1}$; \cite{yam91}).
For this reason, no detection has been reported for the $^{3}${\it{P}}$_{1}$--$^{3}${\it{P}}$_{0}$ line of [$^{13}$C I], as far as we know. 
Since the [C I] line emissions in the 49 Ceti disk are bright and their line widths are narrow, 
the [$^{13}$C I] line emission was fortunately detected for the first time.

Although we obtained high-resolution images of 49~Ceti in the CO and [C I] lines with ALMA, we are facing a difficulty 
that we have little information on the spatially-resolved optical depths at a resolution of 0$\arcsec$5. 
It is thus difficult to derive the spatially-resolved column densities of CO and C from the observed intensities based on the current datasets. 
At this moment, high-quality and high-resolution images are available only for CO($J$=3--2) and [C I] $^{3}${\it{P}}$_{1}$--$^{3}${\it{P}}$_{0}$.
To overcome the above difficulty, we need spatially-resolved images of the other lines of $^{12}$CO and [C I],
as well as those of $^{13}$CO and [$^{13}$C I] lines.
Such observations are strongly awaited to evaluate the C/CO ratio accurately, which will be a key to understand the origin of the gas in 
the debris disk of 49 Ceti.

\bigskip
\acknowledgments
We thank the referee for the thoughtful and constructive comments.
This {\it{Letter}} makes use of the following ALMA data:ADS/JAO.ALMA$\#$2017.0.00467.S.
ALMA is a partnership of ESO (representing its member states), NSF (USA) and NINS (Japan), 
together with NRC (Canada), NSC and ASIAA (Taiwan), 
and KASI (Republic of Korea), in cooperation with the Republic of Chile. 
The Joint ALMA Observatory is operated by ESO, AUI/NRAO and NAOJ.
This study is supported by KAKENHI (18K03713, 19H05090, 19H05069, and 19K14753).
Data analyses were carried out on a common-use data analysis computer system at the Astronomy Data Center, 
ADC, of the National Astronomical Observatory of Japan.

\facility{ALMA}
\software{CASA 5.5.0}

\begin{deluxetable}{l l l l l c c c c}
\tabletypesize{\small}
\label{tb1}
\tablecaption{Line parameters of [C I] and [$^{13}$C I]}
\tablewidth{0pt}
\tablehead{
\colhead{Species} & \colhead{Transition} & \colhead{} & \colhead{Frequency} & \colhead{S$\mu^{2}$\tablenotemark{a}} & \colhead{$V_{\rm{shift}}$\tablenotemark{b}}\\
\colhead{} & \colhead{} & \colhead{} & \colhead{[GHz]}& \colhead{[Debye$^{2}$]} & \colhead{[km~s$^{-1}$]}}
\startdata
\lbrack{C I}\rbrack & $^{3}${\it{P}}$_{1}$--$^{3}${\it{P}}$_{0}$ & -- & 492.160651 & 1.7283$\times$10$^{-4}$ & 0 \\
\lbrack{$^{13}$C I}\rbrack & $^{3}${\it{P}}$_{1}$--$^{3}${\it{P}}$_{0}$ & $F$=3/2--1/2  & 492.164276 &  2.3044$\times$10$^{-4}$ & $-$2.2  \\
\lbrack{$^{13}$C I}\rbrack & $^{3}${\it{P}}$_{1}$--$^{3}${\it{P}}$_{0}$ & $F$=1/2--1/2  & 492.160147 &  1.1523$\times$10$^{-4}$ & 0.3 \\
\hline
\enddata
\tablenotetext{a}{Intrinsic line strength.}
\tablenotetext{b}{The velocity difference from the [C I] emission.}
\end{deluxetable}

\begin{deluxetable}{l l l l l l l l l l l l l l c c c c c c c c c c c c c}
\tabletypesize{\small}
\label{tb2}
\tablecaption{Line parameters of the $F$=3/2--1/2 component of the [$^{13}$C I] $^{3}${\it{P}}$_{1}$--$^{3}${\it{P}}$_{0}$ line}
\tablewidth{0pt}
\tablehead{
\colhead{Position} & \colhead{${T_{\rm{B}}}$} & \colhead{$V_{\rm{LSR}}$\tablenotemark{c}} &  \colhead{$dv$} & \colhead{$\int{{T_{\rm{B}}}}dv$} \\
& \colhead{[K]} & \colhead{[km~s$^{-1}$]} & \colhead{[km~s$^{-1}$]} & \colhead{[K~km~s$^{-1}$]} }
\startdata
(d) & 0.55 (0.11) & 1.6 (0.1) & 0.28 (0.10) & 0.15 (0.08)   \\
(e) & 0.33 (0.11) & 1.6 (0.1) & 0.39 (0.14) & 0.13 (0.09)   \\
(f) & 0.32 (0.11) & 1.9 (0.1) & 0.49 (0.16) & 0.16 (0.10)   \\
\hline
\enddata
\tablenotetext{c}{$V_{\rm{LSR}}$ for the [$^{13}$C I] emission.}
\tablecomments{The numbers in parentheses represent the 1$\sigma$ error.}
\end{deluxetable}

\end{document}